\documentclass[twocolumn,secnumarabic,amssymb, nobibnotes, aps, pra, floatfix,superscriptaddress]{revtex4-1}

\usepackage{graphicx}
\usepackage{amsmath}
\usepackage{amsfonts}
\usepackage{amssymb}
\usepackage{hyperref}
\usepackage[usenames,dvipsnames]{color}

\newcommand{\ket}[1]{\ensuremath{\left| #1 \right \rangle}}
\newcommand{\bra}[1]{\ensuremath{\left\langle #1 \right\rvert}}
\newcommand{\unit}[1]{\ensuremath{\,\mathrm{#1}}}

\newcommand{\expect}[1]{\ensuremath{\langle #1 \rangle}}
\newcommand{\myone}{\leavevmode\hbox{\small1\normalsize\kern-.33em1}}

\begin{document}

\title{Conclusive quantum steering with superconducting transition edge sensors}

\author{Devin H. Smith}\email{smith@physics.uq.edu.au} 
\author{Geoff Gillett}
\author{Marcelo P. de Almeida}
\affiliation{Centre for Engineered Quantum Systems and Centre for Quantum Computation and Communication Technology, University of Queensland, Brisbane, QLD 4072, Australia}
\affiliation{School of Mathematics and Physics, University of Queensland, Brisbane, QLD 4072, Australia}
\author{Cyril Branciard}
\affiliation{School of Mathematics and Physics, University of Queensland, Brisbane, QLD 4072, Australia}
\author{Alessandro Fedrizzi}
\author{Till J. Weinhold}\affiliation{Centre for Engineered Quantum Systems and Centre for Quantum Computation and Communication Technology, University of Queensland, Brisbane, QLD 4072, Australia}
\affiliation{School of Mathematics and Physics, University of Queensland, Brisbane, QLD 4072, Australia}
\author{Adriana Lita}
\author{Brice Calkins}
\author{Thomas Gerrits}
\affiliation{National Institute of Standards and Technology, 325 Broadway, Boulder, CO 80305, USA}
\author{Howard M. Wiseman}
\affiliation{Centre for Quantum Computation and Communication Technology, Centre for Quantum Dynamics, Griffith University, Brisbane, QLD 4111, Australia}
\author{Sae Woo Nam}
\affiliation{National Institute of Standards and Technology, 325 Broadway, Boulder CO 80305, USA}
\author{Andrew G. White}
\affiliation{Centre for Engineered Quantum Systems and Centre for Quantum Computation and Communication Technology, University of Queensland, Brisbane, QLD 4072, Australia}
\affiliation{School of Mathematics and Physics, University of Queensland, Brisbane, QLD 4072, Australia}

\begin{abstract}
Quantum steering allows two parties to verify shared entanglement even if one measurement device is untrusted. A conclusive demonstration of 
steering through the violation of a steering inequality is of considerable fundamental interest and opens up applications in 
quantum communication. To date all experimental tests with single photon states have relied on post-selection, allowing untrusted devices to cheat by hiding unfavourable events in losses. Here we close this ``detection loophole'' by combining a highly efficient source of entangled photon pairs with superconducting transition edge sensors. We achieve an unprecedented $\sim$62\% conditional detection efficiency of entangled photons and violate a steering inequality with the minimal number of measurement settings by 48 standard deviations. Our results provide a clear path to practical applications of steering and to a photonic loophole-free Bell test.
\end{abstract}

\maketitle

Quantum entanglement enables unconditionally secure communication and powerful devices such as quantum computers. In their strongest form, the correlations associated with entanglement rule out locally causal world views in Bell tests~\cite{bell_book}. In a weaker regime, quantum correlations can still be harnessed to \emph{steer} quantum states, demonstrating that two parties, Alice and Bob, share entanglement even if one party is untrusted~\cite{steering_PRL_07}.
Quantum steering was originally introduced by Erwin Schr\"odinger~\cite{schroedinger1935gsq}, in reaction to the Einstein, Podolsky and Rosen (EPR) ``paradox''~\cite{EPR}; it describes the ability to remotely prepare different ensembles of quantum states by performing measurements on one particle of an entangled pair, demonstrating the paradox. Depending on the measurement and its random outcome, the remote system is prepared in a different state; however, the unconditioned remote state remains unaffected, thus preventing any possible superluminal signalling. Interestingly, steering and Bell nonlocality were recently found to be inextricably linked with Heisenberg's uncertainty principle~\cite{oppenheim_Science10}.

The original form of the EPR paradox was demonstrated experimentally with atomic ensembles~\cite{hald1999ssa}, continuous-variable states~\cite{silberhorn2001gcv,bowen2003eic}, and position-momentum entangled single photons\cite{howell2004rep}. More recently, quantum steering was redefined in a quantum information context in~\cite{steering_PRL_07}, promising new applications such as quantum communication using untrusted devices~\cite{CyrilHoward}. 
This new formalisation also allows a strict comparison between the concepts of Bell nonlocality, steering and entanglement~\cite{steering_PRL_07}. In analogy to entanglement witnesses~\cite{Horodeckis96} and Bell inequalities~\cite{bell64}, one can derive experimental criteria~\cite{Cavalcanti_PRA_09} to demonstrate steering: \emph{steering inequalities} impose limits on the observable correlations that can be explained \emph{without} the need to invoke quantum steering. An experimental violation of a steering inequality was recently reported in~\cite{Saunders_NatPhys_10}.

\begin{figure}
	\begin{center}
		\includegraphics[width=.9\columnwidth]{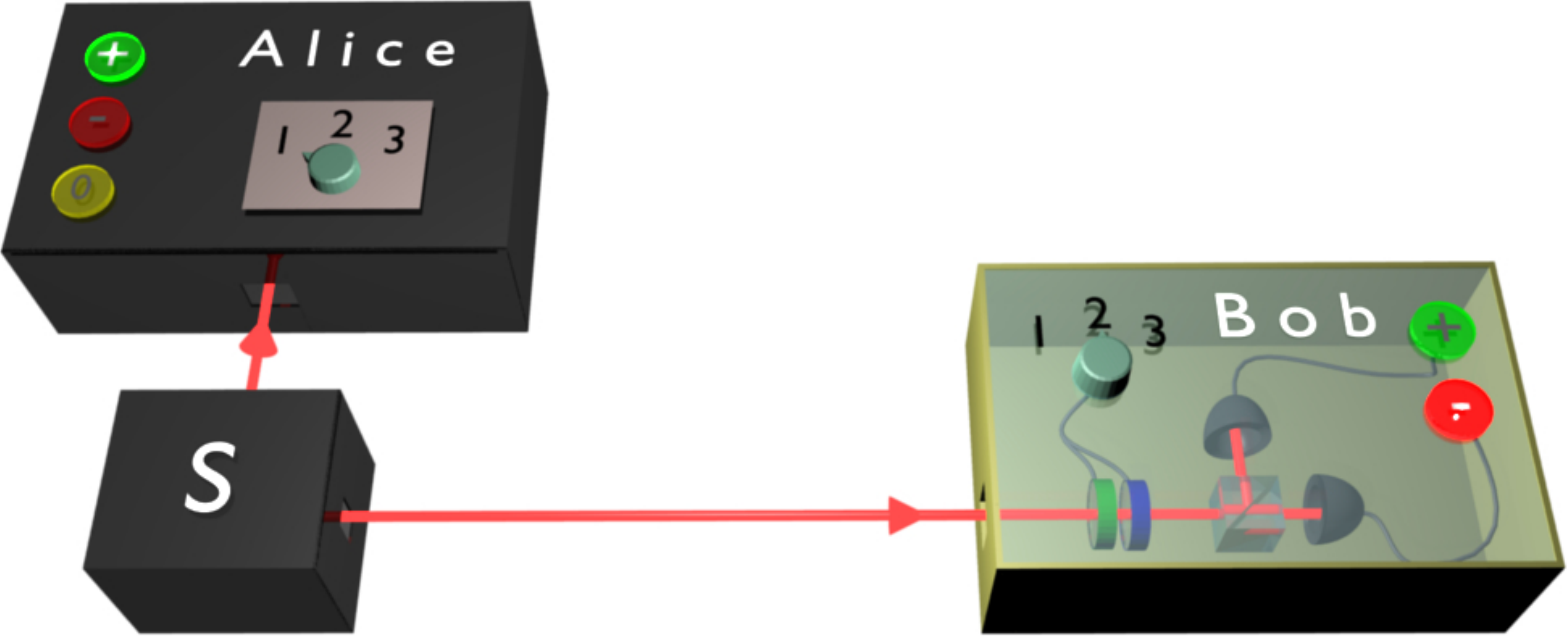}	
	\end{center}
		\caption{\emph{Steering Gedankenexperiment}. Alice and Bob receive particles from a black box (the source, $S$) and  want to establish whether these are entangled. From a prearranged set, they each choose measurements to be performed on their respective particles. Bob's measurement implementation is trusted, but this need not be the case for Alice's; her measurement device is also treated as a black box from which she gets either a ``conclusive'', $A_i = \pm1$, or a ``non-conclusive'' outcome, $A_i = 0$. To demonstrate entanglement, Alice and Bob need to show that she can {\emph{steer}} his state by her choice of measurement.  They can do so through the violation of a steering inequality: whenever Bob's apparatus detects a particle, Alice needs to provide her measurement result. If the recorded correlations of their measurement results surpass the bound imposed by the steering inequality, Alice and Bob have conclusively proven the entanglement of their particles. }
			\label{fig:theory}
\end{figure}

We define the steering task as depicted in Fig.~\ref{fig:theory}: Alice and Bob receive quantum states from a source  and measure them using randomly chosen measurements from a prearranged set; if the observed correlations  violate a steering inequality, then Alice and Bob will be convinced that their shared states were entangled. This holds true even if Alice and Bob trust neither the source nor Alice's measurement device.

Similarly to the case of Bell inequalities, a conclusive violation of a steering inequality requires that the experiment does not suffer from any relevant loopholes. 
When one party has untrusted equipment, the so-called detection loophole~\cite{pearle_DetLH} in particular is critical: if Alice and Bob have to post-select their data on Alice's detected events, then low efficiencies enable her measurement devices to cheat by dropping unfavourable results---in the context of quantum key distribution, for instance,  this would allow the untrusted supplier of the devices to access the key.  The fair sampling assumption invoked in~\cite{Saunders_NatPhys_10} for instance is not satisfactory for such untrusted devices. 

Here we close the detection loophole by using an entangled photon source with high pair collection efficiency~\cite{kim2006pss,Fedrizzi:2007ys} and highly-efficient transition edge sensors~\cite{Lita:2008uq}. We test a steering inequality which naturally accounts for Alice's non-detected events, and violate it by at least 48 standard deviations with the minimum number of two measurement settings
and by more than 200 standard deviations for measurements in three different bases.

\section*{Results}
\subsection*{A quadratic steering inequality for qubits} 
To demonstrate steering, Alice and Bob need to be able to freely choose and perform different measurements; we consider the case where each of them can perform $N = 2$ or 3 measurements, labeled $i,j = 1,2$ or $3$, with \textit{a priori} binary outcomes $A_i,B_j = \pm 1$, see Fig.~\ref{fig:theory}.
Since Bob trusts his measuring device his measurement can be described by a well characterised quantum observable $\hat{B}_j$. He considers only the cases where his measurement gives him a conclusive result, that is, when at least one of his detectors clicks --- if both click, Bob outputs a random result. In contrast, Alice's devices are not trusted and her measurement apparatus is considered a black box. It returns outcomes $A_i = \pm 1$, indicating conclusive measurement results, or $A_i = 0$ when no or both detectors fire. Because Alice must output a result whenever Bob registers an event, these inconclusive results cannot be discarded from further analysis.

The correlation observed by Alice and Bob can be described by the probability distribution $P(A_i=a,B_j=b)$, with $a =\pm 1$ or $0$, and $b = \pm 1$. If Bob receives a state that is not entangled to Alice's the set of possible correlations will be restricted, as the inequality below shows.
First, we define Bob's expectation value for a measurement conditioned on Alice's result:
\begin{equation}
\expect{\hat{B}_i}_{A_i = a} \equiv \ P(B_i \! = \! + 1|A_i \! = \! a) \, - \, P(B_i \! = \! - 1|A_i \! = \! a). \nonumber 
\end{equation}
Averaging this over Alice's results, we define
\begin{equation}
E \big[ \expect{\hat{B}_i}_{A_i}^2 \big] \ \equiv \ \sum_{a = \pm 1, 0} P(A_i = a) \ \expect{\hat{B}_i}_{A_i = a}^2 \,. \label{eq_def_E}
\end{equation}
As shown in the Methods section, if the correlation $P$ could be explained by the source sending Bob an unentangled two-level system (qubit)---that is, in the terminology of~\cite{steering_PRL_07}, if the correlation admits a ``local hidden state'' model---and if Bob implements qubit measurements in two or three mutually unbiased bases, for instance the Pauli $\hat{X}$, $\hat{Y}$ and $\hat{Z}$ operators, then the following inequality holds:
\begin{equation}
S_N \ \equiv \ \sum_{i = 1}^N \ E \big[ \expect{\hat{B}_i}_{A_i}^2 \big] \ \leq \ 1 . \label{steering_ineq}
\end{equation}
Note that the upper bound above depends crucially on Bob's measurement settings, which in experimental implementation will not be perfectly orthogonal, nor perfectly projective; we detail in the Methods section how the bound must be corrected to account for experimental imperfections.

Quantum mechanics allows a violation of inequality~(\ref{steering_ineq}), which thus implies steering. To get a first insight, suppose that Alice and Bob share Werner states 
of visibility $V$, $\rho =V\ket{\psi^{-}}\bra{\psi^{-}}+(1-V)\openone/4$, where $\ket{\psi^{-}}$ is the Bell singlet state, and that Alice implements the same measurements as Bob. Then, due to the anti-correlation of the singlet state, $\expect{\hat{B}_i}_{A_i = \pm1} = \mp V$: this illustrates that Alice can steer Bob's state to be aligned with her measurement axis, limited by the visibility of the shared state.
If Alice has a probability $\eta$ of getting a conclusive outcome whenever Bob gets one, then $E \big[ \expect{\hat{B}_i}_{A_i}^2 \big] = \eta V^2$. This implies that the steering inequality \eqref{steering_ineq} will be violated if
\begin{equation}
\eta V^2 > 1/N . \label{visibility-eq}
\end{equation}

Satisfying the requirements given by Eq. \eqref{visibility-eq} in a photonic architecture is challenging. 
For the minimal set of $N{=}2$ measurement settings, an experimental test of the steering inequality \eqref{steering_ineq} requires, even for a pure entangled singlet state with visibility $V{=}1$, that Alice detects a signal more than $\eta{>}50\%$ of the times Bob requests a response. To reach these requirements, the experimental apparatus has to be carefully optimised.

\subsection*{Experimental setup}
\begin{figure}[t]
	\begin{center}
		\includegraphics[width=\columnwidth]{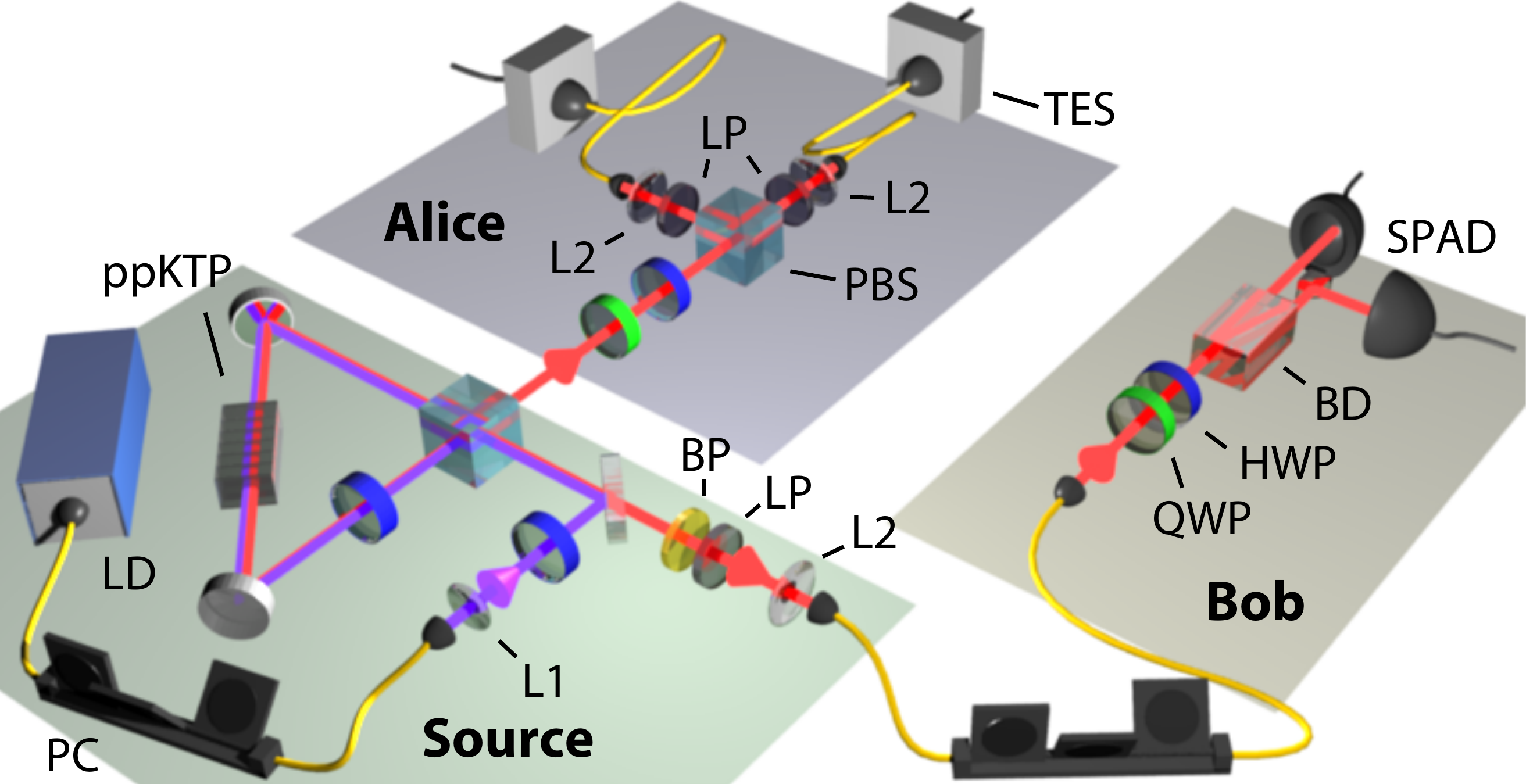}	
	\end{center}
	\caption{\emph{Experimental scheme}. Polarisation-entangled two-photon states are generated in a periodically poled 10\unit{mm} KTiOPO$_{4}$ (ppKTP) crystal inside a polarisation Sagnac loop ~\cite{kim2006pss,Fedrizzi:2007ys}. The continuous wave, grating-stabilised 410\unit{nm} pump laser (LD) is focussed into this crystal with an aspheric lens (L1, f=4.0\unit{mm}) and its polarisation is set with a fibre polarisation controller (PC) and a half-wave plate (HWP), controlling the entangled output state~\cite{kim2006pss}. Bob filters his output photon with a long-pass glass filter (LP) and a 3\unit{nm} band-pass filter (BP), before collecting it with an aspheric lens (L2, f=18.4\unit{mm}) into a single-mode fibre.
	  He performs his measurement in an external fibre bridge, with a combination of a quarter-wave plate (QWP), HWP, a polarising beam displacer (BD) and multi-mode-fibre-coupled single-photon avalanche diodes (SPADs). To minimise loss, Alice performs her measurement directly at the source using a QWP, HWP and a polarising beamsplitter (PBS), followed by a LP filter and fibre collection with focussing optics identical to Bob's, finally detecting her photons with highly efficient, superconducting transition edge sensors (TESs)~\cite{Lita:2008uq}.}
	\label{fig:setup}
\end{figure}

We performed our experiment using entangled photons created in a polarisation Sagnac source based on spontaneous parametric downconversion~\cite{kim2006pss,Fedrizzi:2007ys},  see Fig.~\ref{fig:setup}. This source design meets two crucial requirements; a high entangled-pair collection efficiency and near-ideal polarisation entanglement.

We followed~\cite{Fedrizzi:2007ys} in the basic design of our source.  To maximise the conditional coupling between the Alice and Bob's collection apparatus,   we optimised the pump and collection spots based on~\cite{bennink2010ocg},  with the optimum found at using  pump spot and collection mode diameters of $200\unit{\mu m}$ and $84\unit{\mu m}$ in the crystal, respectively. With these parameters, we achieved typical pair detection efficiency of 40\% measured with standard single-photon avalanche diodes (SPADs), whose detection efficiency was estimated to be 50\% at $820\unit{nm}$, implying a collection efficiency of 80\%. Due to the asymmetry of the steering task, the source and detection system do not have to be symmetric. For example, in our setup Alice does not employ narrow-band filters; this choice increases her overall background, but reduces her loss, thus increasing the detection efficiency conditioned on Bob's measurement.

Another key requirement is high photon detection efficiency. The upper bound of the conditional detection probability $\eta$ is limited by the performance of Alice's photon detectors, and therefore would not even in a loss-less, noise-free case allow us to meet the requirements of eq. \eqref{visibility-eq} with our SPADs and two measurement settings; in our experiment, Alice thus employs superconducting tungsten transition edge sensors~\cite{Lita:2008uq} (TESs). TESs utilise a layer of superconducting tungsten kept in the transition temperature range and offer a combination of photon number resolution and high detection efficiency of up to $95$\% at $1550$\unit{nm}, while being virtually free of dark counts~\cite{Lita:2008uq}. Our detectors were optimised for 810\unit{nm} with an optical cavity similar to that presented in an earlier work~\cite{Lita:2008uq}, with an estimated detection efficiency for $820\unit{nm}$ photon in the $1550\unit{nm}$ single-mode SMF-28 fibre connected to this cavity to be larger than 97\%. In practice, the measured detection efficiencies of our two TES  were $1.50$ and $1.56$ times higher than the efficiency of our reference SPAD at 820\unit{nm}.The dominant source of optical loss, which leads to these less-than-optimal figures, was a splice between the  single mode 820\unit{nm} fibres connected to the source and the fibres connected on the TES, which were single mode at 1550\unit{nm}.

The TESs were operated between 40\unit{mK} and 75\unit{mK} and yielded analog output pulses with a rise time of $\sim$320\unit{ns} and jitter of ${\sim}78$\unit{ns}.
In order to detect coincidences between the TES signals and the TTL pulses generated by the SPADs each amplified TES signal was digitised with a constant fraction discriminator; because the TESs rethermalise after each detection event, with a relaxation period of ${\sim}2\unit{\mu s}$, the non-number resolving discriminators were set to impose a dead time of the same length to avoid  false detections during the TES relaxation period. To match the delay caused by the TES detection system, Bob's SPAD signals were delayed by ${\sim}450$\unit{ns}. Coincident events were then detected with a field-programmable gate array with a timing window of 98\unit{ns}. 

The long dead time period imposed by our electronics leads to a rate-dependent loss and we therefore operated the source at comparatively low rates of photon-pair creation. We achieved optimal conditional detection efficiency at a laser pump power of 250\unit{\mu W}, generating ${\sim}12$\unit{kHz} of single photons in each TES channel. At this rate, the loss due to dead time was ${\sim}2.5\%$.

\subsection*{Experimental violation of our steering inequality}
\begin{figure}[t]
	\begin{center}
		\includegraphics[width=\columnwidth]{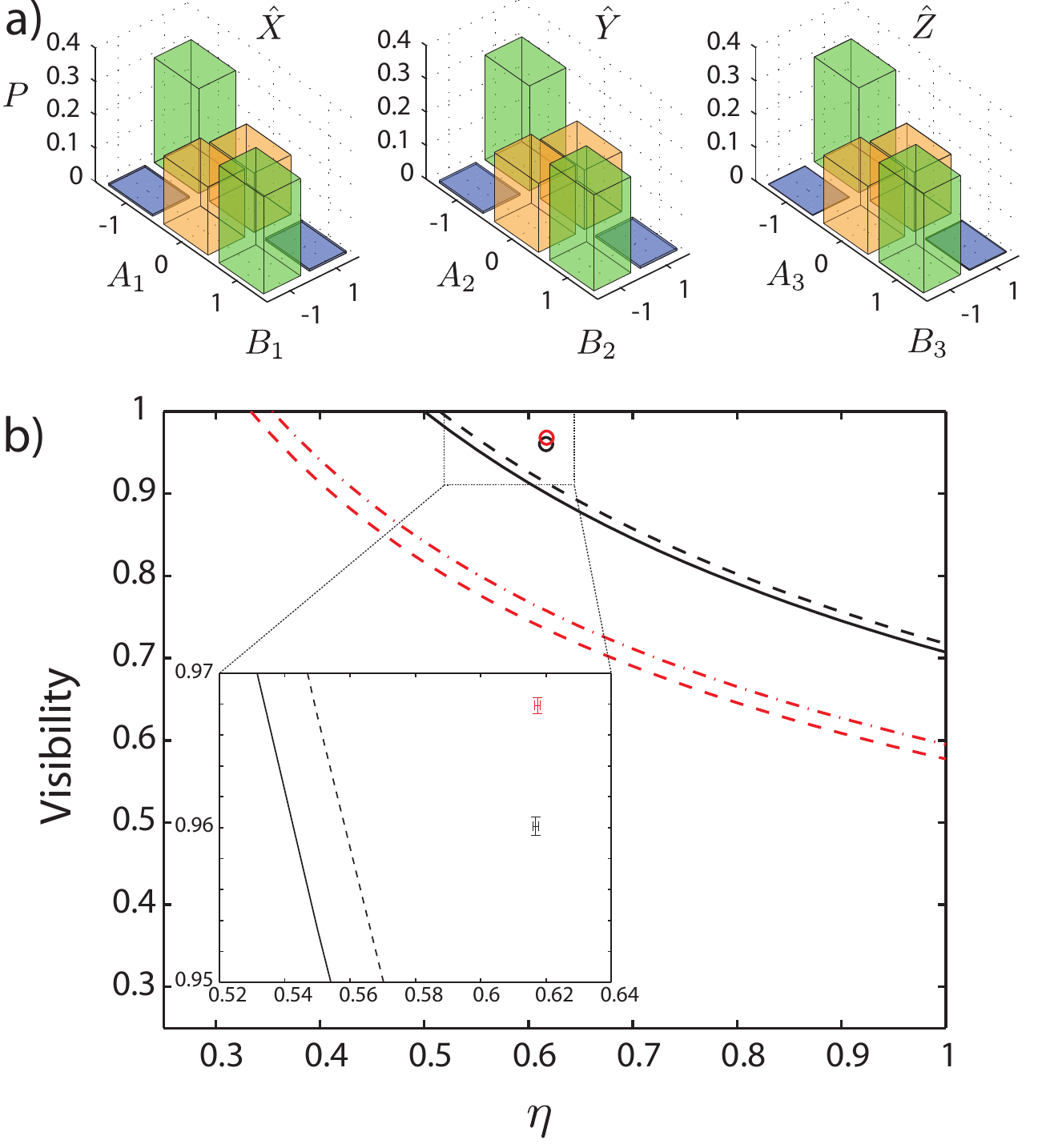}	
	\end{center}
	\caption{\emph{Experimental results}. (a) Probability distributions $P(A_{i}{=}a,B_{i}{=}b)$ for the $S_{3}$ measurements $\hat{X}$, $\hat{Y}$ and $\hat{Z}$, calculated by normalising the registered coincident events for each measurement setting to the total count numbers. The green and blue bars represent correlations that indicate the quality of the shared entangled state. The orange bars represent events that Alice failed to detect. Error bars are too small to be seen on this scale. For $S_{2}$, we used the data obtained from the measurements of $\hat{X}$ and $\hat{Y}$. (b) Theoretical visibility required to violate steering inequalities for $N{=}3$ (red dashed line) and $N{=}2$ (black line) for a given efficiency $\eta$. Our measurement clearly violates this bound, with an averaged visibility of $V{=}0.9678\pm0.0005$ at a mean heralding efficiency of $\eta{=}0.6175\pm0.0008$ for the $N{=}3$ measurements (red) and $V{=}0.9601 \pm0.0006$, $\eta{=}0.6169\pm0.0008$ for $N{=}2$ (black). 
 All error bars (one standard deviation) were calculated assuming Poissonian photon-counting statistics. The correction of the analytic bounds of \eqref{fig:viseta} due to measurement imprecision (see the Methods section) is shown by the dash-dotted red line for $S_{3}$ and the dashed black line for $S_{2}$.}
	\label{fig:viseta}
\end{figure}

We produced the polarisation-entangled singlet state $\ket{\psi^{-}}=(\ket{HV}-\ket{VH})/\sqrt{2}$, where $\ket{H}$ and $\ket{V}$ represent single horizontal and vertical polarised photons respectively, and performed separate measurements in 2 and 3 different bases ($N{=}2,3$, with measurements of $\hat{X}$, $\hat{Y}$ and $\hat{Z}$). The discrete probability distribution for Alice and Bob's correlations, $P(A_i{=}a,B_j{=}b)$, is shown in Fig.~\ref{fig:viseta}~(a). From these, we first estimated an averaged heralding efficiency $\eta$ and entangled state visibility $V$ and compared them to the theoretical minimum requirements, Eq.~(\ref{visibility-eq}), in Fig.~\ref{fig:viseta}~(b). The plot indicates that we should expect a conclusive, detection-loophole-free demonstration of steering.

Indeed, for the steering parameter $S_3$ defined in~(\ref{steering_ineq}), we obtained
\begin{equation}
S_3=1.7408\pm0.0017\,, \nonumber
\end{equation}
where the uncertainty (one standard deviation) was calculated by standard propagation of the Poissonian photon-counting statistics. The corrected bound, due to imprecision in Bob's measurements and as calculated in the Methods section, was $1.062\pm0.003$. This corresponds to a violation of the inequality \eqref{steering_ineq} by more than 200 standard deviations. 

For $N=2$, the corresponding corrected bound of inequality~\eqref{steering_ineq} was $1.0291\pm0.0019$. We obtained the value
\begin{equation}
S_2=1.1410 \pm 0.0014  \nonumber
\end{equation}
for the experimental steering parameter, yielding a violation of the steering inequality by $48$ standard deviations.

\section*{Discussion}

Our highly efficient system allows us to firmly close the detection loophole in our demonstration of quantum steering, achieving the highest ever reported heralding efficiency for entangled photons, $\eta\sim62\%$. Our experimental violation of inequality \eqref{steering_ineq} has a quite intuitive interpretation: it shows that Alice can, at her will, \emph{steer} Bob's qubit state to be preferably polarised along any of the three axes of the Bloch sphere, see Fig.~\ref{fig:viseta}~(a).

While we have closed the detection loophole, we have not addressed the locality  and freedom of choice loopholes\cite{scheidl2010vlr} in this work; closing these would require Alice and Bob's choice and implementation of measurements to be space-like separated, as demonstrated in a very recent experiment reported in~\cite{Vienna_paper}. For practical purposes in quantum communication, however, these loopholes are typically not problematic\cite{Pironio_DIQKD_NJP}: it is a necessary assumption that Alice and Bob can choose their measurements independently of the state preparation, and that no unwanted information leaks from Alice and Bob's laboratories.

Besides the criteria employed here there are others that can be used to demonstrate steering\cite{Cavalcanti_PRA_09,walborn11}. If Alice cannot achieve the high heralding efficiencies obtained in our experiment, some of these may be advantageous: as recently shown in~\cite{GU_paper}, generalising the linear criteria of~\cite{Saunders_NatPhys_10} allows for steering with arbitrarily high losses. These however require a larger number of different measurement settings; the experiment reported in~\cite{GU_paper} used up to $N{=}16$ measurements. 
Our choice to test inequality~(\ref{steering_ineq}) was motivated by its simplicity in how it naturally accounts for Alice's detection inefficiencies, and by its minimality in the number of settings. Note that $N{=}2$ is the number of settings initially discussed by EPR; it is also the canonical number of settings in applications to quantum cryptography~\cite{CyrilHoward}.

Increasing Alice's detection efficiency above 66\% will infact enable steering to be used for quantum key distribution where one party distrusts their apparatus\cite{CyrilHoward}; our experiment thus constitutes an important step towards practical applications of quantum steering. Furthermore, our results imply that a fully loophole-free photonic Bell test seems to be within arm's reach. While the symmetric photon pair detection efficiency for our setup is somewhat lower than the conditional detection probability $\eta$, it is not far below the $66.\bar{6}$~\% limit required to violate a Clauser-Horne inequality~\cite{ClauserHorne74} with non-maximally entangled states~\cite{eberhard1993blc}. Although still a technological challenge, it is now becoming conceivable to surpass this efficiency in the near future, while simultaneously addressing the locality and freedom-of-choice loopholes such as demonstrated in~\cite{scheidl2010vlr}. 

\section*{Methods}
\subsection*{Proof of inequality~\eqref{steering_ineq}}
Inequality~(\ref{steering_ineq}) is equivalent to previously derived variance criteria~\cite{Cavalcanti_OptExpr_09,Cavalcanti_PRA_09}; for completeness, we give here a simple proof.

If the observed correlation can be explained by the source sending non-entangled states to Alice and Bob, then the probability distribution $P$ can be decomposed in the form
\begin{equation}
P(A_i \!=\! a, B_j \!=\! b) = \sum_{\lambda} q_{\lambda} \, P_{\lambda}(A_i \!=\! a) \, P_{\rho_{\lambda}}^Q(B_j \!=\! b), \nonumber
\end{equation}
where $\lambda$ describes the source preparation, used with probability $q_{\lambda}$ (such that $q_{\lambda} \geq 0$, $\sum_{\lambda} q_{\lambda} = 1$ --- note that the sum could in principle be continuous and infinite): it specifies Alice's response function $P_{\lambda}(A_i \!=\! a)$ implemented by her (untrusted) measurement device, and the state $\rho_{\lambda}$ sent to Bob. Bob's response function $P_{\rho_{\lambda}}^Q(B_j \!=\! b)$ is then as quantum mechanics predicts when the observable $\hat{B}_j$ is measured on $\rho_{\lambda}$.

From the above decomposition, and defining $$q_{\lambda|A_i=a} \equiv q_{\lambda} \, \frac{P_{\lambda}(A_i \!=\! a)}{P(A_i \!=\! a)}, $$ such that, as before, $q_{\lambda|A_i=a} \geq 0$ and $\sum_{\lambda} q_{\lambda|A_i=a} = 1$, we get $P(B_i{=}b | A_i {=} a){=}\sum_{\lambda} q_{\lambda|A_i{=}a} \, P_{\rho_{\lambda}}^{Q}(B_i {=} b)$ and
\begin{equation}
\expect{\hat{B}_i}_{A_i = a} \ = \ \sum_{\lambda} q_{\lambda|A_i=a} \, \expect{\hat{B}_i}_{\rho_{\lambda}} \,. \nonumber
\end{equation}
By the convexity of the square,
\begin{equation}
\expect{\hat{B}_i}_{A_i = a}^2 \ \leq \ \sum_{\lambda} q_{\lambda|A_i=a} \, \expect{\hat{B}_i}_{\rho_{\lambda}}^2 , \nonumber
\end{equation}
which leads to
\begin{equation}
E \big[ \expect{\hat{B}_i}_{A_i}^2 \big] \ \leq \ \sum_{\lambda} q_{\lambda} \, \expect{\hat{B}_i}_{\rho_{\lambda}}^2 . \nonumber 
\end{equation}

Now, for any 1-qubit state $\rho_{\lambda}$, and for 3 mutually unbiased observables $\hat{B}_i$, one has
\begin{equation}
\sum_{i=1}^3 \ \expect{\hat{B}_i}_{\rho_{\lambda}}^2 \ \leq \ 1 \,. \nonumber
\end{equation}
Together with the previous inequality and the normalisation $\sum_{\lambda} q_{\lambda} = 1$, we obtain inequality \eqref{steering_ineq} for $N = 3$; the case for $N = 2$ follows trivially.

\subsection*{Accounting for experimental imperfections in Bob's measurements}

As already highlighted, inequality \eqref{steering_ineq} is highly dependant on Bob's measurements: it is only valid when Bob measures mutually unbiased observables on qubits. In a practical experiment however, Bob will not measure along perfectly mutually unbiased bases, and his operators may not act on a 2-dimensional system only. We show now that the parameter $S_N$ can still be used to demonstrate steering, but the upper bound in \eqref{steering_ineq} must be adapted according to Bob's actual measurement.

Let us start by giving a more accurate description of the measurement Bob performs in our experiment. First, he uses quarter- and half-wave plates, that define a direction (\textit{i.e.}, a unit vector) $\vec b$ on the Bloch sphere, representing his choice of basis. The beam displacer (BD) then separates the $H$ and $V$ polarisations: a fraction $t \simeq 1$ of the $H$ polarisation goes to its first output channel, and later on to the ``+1" detector, while a fraction $1-t \ll 1$ (in our experiment, $1-t \leq 10^{-5}$) goes to the second output channel, and to the ``-1" detector. We can assume that,  symmetrically, a fraction $t$ of the $V$ polarisation goes to the second output channel, while a fraction $1-t$ goes to the first output channel, as we utilise a calcite beam displacer as our polarising element; its intrinsic birefringence maps polarisation into different spatial modes, which is a fundamentally symmetric effect~\cite{Hecht} other polarising elements rely on other effects, not necessarily symmetric, requiring a slightly more thorough analysis. We finally denote by $\eta_+$ and $\eta_-$ the overall detection efficiencies of the +1 and -1 detectors (SPADs), respectively, including all losses in Bob's lab, including coupling and detection losses. 

For a single photon state entering Bob's lab, represented by a vector $\vec u$ in the Bloch sphere, the probability that it gives a click on the +1 or -1 detector is then
\begin{equation}
P_B(\pm|\vec b) = \frac{\eta_{\pm}}{2} \big[ 1 \pm (2t-1) \vec b \cdot \vec u \big] \,. \nonumber
\end{equation}
It follows that
\begin{eqnarray*}
P_B(+|\vec b) + P_B(-|\vec b) & \ = \ & \frac{\eta_{+}\!+\!\eta_{-}}{2} + \frac{\eta_{+}\!-\!\eta_{-}}{2} \, (2t-1) \, \vec b \cdot \vec u \\
&& \hspace{-2cm} \geq \ \frac{\eta_{+}+\eta_{-}}{2} - \Big| \frac{\eta_{+}-\eta_{-}}{2} \Big| \ = \ \min(\eta_+,\eta_-) \\
\end{eqnarray*}
and
\begin{eqnarray*}
|P_B(+|\vec b) - P_B(-|\vec b)| & = & \big| \frac{\eta_{+}\!-\!\eta_{-}}{2} \!+\! \frac{\eta_{+}\!+\!\eta_{-}}{2} \, (2t-1) \, \vec b \cdot \vec u \big| \\
&& \hspace{-2cm} \leq \ \big| \frac{\eta_{+}\!-\!\eta_{-}}{2} \big| + \frac{\eta_{+}\!+\!\eta_{-}}{2} \, \big| \vec b \cdot \vec u \big| \\
&& \hspace{-1cm} = \ \max(\eta_+,\eta_-) \Big[ \delta + (1-\delta) |\vec b \cdot \vec u| \Big]
\end{eqnarray*}
with $\delta \equiv \frac{| \eta_{+}-\eta_{-} |}{2\max(\eta_+,\eta_-)}$.
Hence, defining $w \equiv \frac{\max(\eta_+,\eta_-)}{\min(\eta_+,\eta_-)}$, we get 
\begin{equation}
|\expect{B}| = \frac{|P_B(+|\vec b) - P_B(-|\vec b)|}{P_B(+|\vec b) + P_B(-|\vec b)} \leq w \, \big[ \delta + (1-\delta) |\vec b \cdot \vec u| \big] \nonumber
\end{equation}
and by convexity,
\begin{equation}
\expect{B}^2 \leq w^2 \, \big[ \delta + (1-\delta) (\vec b \cdot \vec u)^2 \big] \,. \nonumber
\end{equation}

Consider now $N = 2$ or 3 measurement directions $\vec b_i$, such that $|\vec b_i \cdot \vec b_j| \leq \epsilon$ for all $i \neq j$, for some $\epsilon > 0$ quantifying the nonorthogonality of the $N$ directions.
One can show that for all $\vec u$ in the Bloch sphere,
\begin{equation}
\sum_{i = 1}^N \ ( \vec b_i \cdot \vec u )^2 \ \leq \ 1 + (N-1) \, \epsilon \nonumber.
\end{equation}
Indeed the worst case is obtained when the $N$ vectors $\vec b_i$ are such that $ \vec b_i \cdot \vec b_j = \epsilon$ for all $i \neq j$, and when $\vec u$ is a unit vector equidistant to the $\vec b_i$, which gives the upper bound above.

Following the proof of inequality~\eqref{steering_ineq}, we now obtain, for Bob's actual measurements, the steering inequality
\begin{equation}
S_N \leq w^2 \big[ 1 + (N-1)(\delta + \epsilon - \delta \epsilon) \big] \, . \label{steering_ineq_asym}
\end{equation}

In our experiments, the ratio of detection efficiencies in Bob's two detectors was found to be $w=\eta_{+}/\eta_{-} = 1.0115\pm0.0007$. We estimated the orthogonality of Bob's measurements (\textit{i.e.}, $\epsilon$) by inserting a large ensemble of different linear polarisation states into Bob's measurement device, fully characterising the two wave-plates and the relative coupling. For the $N{=}3$ measurement settings, we can take epsilon to be the maximum of all three scalar products $|\vec b_i \cdot \vec b_j|$; we found  $\epsilon{=}0.0134{\pm}0.0007$ in that case. For the test with $N{=}2$ settings, we used the two most orthogonal settings, $\hat{X}$ and $\hat{Y}$, which gave $\epsilon{=}(1.3{\pm}1.5){\times}10^{-4}$. From~\eqref{steering_ineq_asym}, this yields bounds of $1.0291\pm0.0019$ and $1.062\pm0.003$ for $S_2$ and $S_3$, respectively, as quoted in the main text.

Other experimental imperfections include dark counts and background of the detectors at tens of hertz, compared to a rate of ${\sim}12$\unit{kHz} total detected events for Bob. These will however only introduce some white noise into Bob's data, and cannot increase the bound in the steering inequality.

Note finally that all the above calculations assumed that Bob received qubit states, encoded in the polarisation of single photons. This is however not guaranteed, and the source could indeed send multiphoton states.  These possibly lead to double clicks; in the experiment, a negligible fraction of about 1 in $10^{4}$ of Bob's events were double-click events. However, we believe that a careful analysis along similar lines as in~\cite{moroder_squashing} should show that if Bob gives a random result when he gets a double click (rather than discarding the event), multiphoton states are then also bound to satisfy the steering inequality \eqref{steering_ineq_asym}.  Indeed double clicks will then reduce Alice and Bob's correlations, and will not help Alice's untrustworthy devices to increase the steering parameter $S_N$. 
\section*{Acknowledgements}

We thank M.A. Broome for assistance with the experimental preparations. We acknowledge financial support from the ARC Discovery and Federation Fellow programs and an IARPA-funded US Army Research Office contract. This research was conducted by the Australian Research Council Centres of Excellence for Engineered Quantum Systems (Project number CE110001013) and Quantum Computation and Communication Technology (Project number CE110001027). The NIST contribution was supported by the NIST Quantum Information Initiative, and is a work of the US Government and as such this article is not subject to US Copyright.

\bibliographystyle{naturemag}

\end{document}